\documentclass[prc,twocolumn,showpacs,amsmath,amssymb,superscriptaddress,floatfix,nofootinbib]{revtex4}
\usepackage{mathrsfs,bm}
\usepackage{longtable,lscape}
\usepackage{txfonts}
\usepackage{amssymb}
\usepackage{indentfirst}
\usepackage{graphicx,,booktabs}
\usepackage{multirow}
\usepackage{color}
\usepackage{amssymb}

\begin{document}
\title{Theoretical investigation of the decay of the $N(2120)$ resonance to nucleon resonances near 1.7 GeV}

\author{Yin Huang}
\email{huangy2014@lzu.cn}\affiliation{Research Center for Hadron and CSR Physics, Lanzhou University and Institute
of Modern Physics of CAS , Lanzhou 730000,
China}\affiliation{School of Nuclear Science and Technology,
Lanzhou University, Lanzhou 730000, China}
\affiliation{Institute of Modern Physics,
Chinese Academy of Sciences, Lanzhou 730000, China}
\author{Jun He\footnote{Corresponding
author}} \email{junhe@impcas.ac.cn}\affiliation{Research Center for Hadron and CSR Physics, Lanzhou University and Institute
of Modern Physics of CAS , Lanzhou 730000,
China}
\affiliation{Institute of Modern Physics, Chinese Academy of
Sciences, Lanzhou 730000, China}  \affiliation{State Key
Laboratory of Theoretical Physics, Institute of Theoretical Physics,
Chinese Academy of Sciences, Beijing 100190, China}
\author{Xu-Rong Chen}
\affiliation{Research Center for Hadron and CSR Physics, Lanzhou University and Institute of Modern Physics of CAS , Lanzhou 730000,
China}\affiliation{Institute of Modern Physics, Chinese Academy of
Sciences, Lanzhou 730000, China}
\author{Rong Wang}
 \affiliation{Research Center for Hadron and CSR Physics, Lanzhou University and Institute
of Modern Physics of CAS , Lanzhou 730000,
China}\affiliation{School of Nuclear Science and Technology,
Lanzhou University, Lanzhou 730000, China}\affiliation{Institute of Modern Physics, Chinese Academy of
Sciences, Lanzhou 730000, China}

\author{Ju-Jun Xie }
\affiliation{Research Center for Hadron and CSR Physics, Lanzhou University and Institute
of Modern Physics of CAS , Lanzhou 730000,
China} \affiliation{Institute of Modern Physics,
Chinese Academy of Sciences, Lanzhou 730000, China}\affiliation{State Key
Laboratory of Theoretical Physics, Institute of Theoretical Physics,
Chinese Academy of Sciences, Beijing 100190, China}
\author{Hong-Fei Zhang}
  \affiliation{School of Nuclear
Science and Technology, Lanzhou University, Lanzhou 730000, China}\affiliation{Institute of Modern Physics, Chinese Academy of
Sciences, Lanzhou 730000, China}

\date{\today}
\begin{abstract}\begin{description}
\item[Background:] Until now the knowledge about nucleon resonances
	with a mass higher than 2 GeV  has been scarce. Huge amounts of experimental data of the multipion photoproduction have been accumulated, and more can be expected in the future in facilities such as JLab 12 GeV. It makes it possible to investigate the decay of a nucleon resonance into another nucleon resonance.
\item[Purpose:] The possibility to research the decay of the $N(2120)$ to
	nucleon resonances near 1.7 GeV in the three-pion photoproduction will be explored to provide useful information for future experimental study.
\item[Method:] The pion and radiative decay amplitudes of nucleon resonances are studied within the constituent
	quark model, which is used to calculate the couplings constants, especially for the decay of a nucleon resonance near 2.1 GeV to another nucleon resonance near 1.7 GeV.
	The three-pion photoproduction off the neutron target, i.e.,$\gamma n \to\pi^{-}\pi^{-}\Delta^{++}\to \pi^{-}\pi^{-}\pi^+ p$,
	 is investigated based on the effective Lagrangian method with the coupling constant obtained form the decay amplitudes.
\item[Results:]  The  resonance contribution with a state $N(^2P_M)\frac{3}{2}^{-}$ near 2.1 GeV decaying to a state
	$N(^4P_M)\frac{5}{2}^{-}$ near 1.7 GeV, i.\ e.,  $N(2120)\to N(1675)\pi$ is dominant in the process considered.
	The total cross section from the resonance contribution is at the order of 1 $\mu$b and can be easily distinguished from the background.
\item[Conclusions:] Our results suggest it is practicable to research the decay of the $N(2120)$
	to the $N(1675)$ in experiment.
\end{description}
\end{abstract}


\maketitle
\section{INTRODUCTION}

With improvement of experimental technique, the understanding about
nucleon resonances has deepened in recent years. More nucleon
resonances near 1.9 GeV were listed in the new version of the review of
particle physics [Particle Data Group (PDG)]~\cite{prd9}, which is a great progress about
the long standing ``missing resonances'' problem.  However, the
knowledge about resonances with mass higher than 2 GeV are still scarce.  The nucleon resonances observed is much less than those
predicted in the constituent quark model. For the nucleon resonances
observed, the internal structure is still in controversy. For example,
based on the prediction in the constituent quark
model~\cite{Capstick:1998uh,Capstick:1992uc}, a nucleon resonance
near 2.1 GeV, which is the third $N3/2^-$ state, should play an important role in the $\Lambda(1520)$
photoproduction\cite{He:2014gga,He:2012ud}. However, Klempt claimed
that the $N(1875)$ instead of $N(2120)$ is the missing third $N3/2^-$
state~\cite{Klempt:2009pi}.

It was predicted in the constituent quark model that more nucleon resonances will be found in the higher mass region.  For example, the number of $N=2$ shell
states is much larger than that of $N=1$ shell states.  It requires experimental data from more channels  to distinguish the
predicted nucleon resonances above 2 GeV.   Currently most information
about nucleon resonances is extracted from pion nucleon scattering and
single meson photoproduction.   A high-mass nucleon resonance is
prone to decay into a baryon resonance with a meson, which provides a
new way to detect internal structure of  high-mass nucleon resonance.
In recent years, a few attentions have been attracted by the decay of
a nucleon resonances to a baryon resonance. For example, due to the
high thresholds of the $\Lambda(1520)$ and $\Sigma(1385)$
photoproductions it was suggested by several authors that the new
experimental data released by the CLAS Collaboration
\cite{Moriya:2013hwg} are very useful to  study  a nucleon resonance
near 2.1 GeV~\cite{He:2014gga,He:2013ksa,Xie:2013mua}.  In order to
analyze various isospin channels of double-pion production in
nucleon-nucleon collisions, the contribution from $\Delta(1600)$
decaying to the Roper resonance $N(1440)$ in process $NN\to{}NN\pi\pi$ was
mentioned~\cite{prd99}. However, up to now only the branch ratio of
$\Delta(1600)\to\pi{}N(1440)$ has been provided by the PDG~\cite{prd9}.

The two-pion and three-pion photoproductions have been
measured in many facilities, such as CLAS@JLab and MAMI, and a large
amount of data have been accumulated in recent years. Most
analyses about these data aimed to study the light meson spectrum and to
search for the exotic meson. Besides the light meson production with
$t$ channel exchange mechanism,  if the photon carries enough energy the initial nucleon will be  excited to a nucleon resonance with a mass higher than 2 GeV,  then decay to a baryon resonance with lower mass.  For example, the
$\Lambda(1520)$ and $\Sigma(1385)$ photoproductions are extracted from
multimeson photoproduction with nucleon \cite{Moriya:2013hwg}. The high energy and high intensity photon beam in the CLAS@JLab experiment
provides an opportunity to study nucleon resonance in the high-mass range.
In the future experiments at GlueX and CLAS12 detectors, more data
will be collected, which makes it more feasible to study the decay of a nucleon resonance into another resonance.

One reaction of interest is $\gamma n \to \pi^- \pi^-
\Delta^{++}\to\pi^-\pi^-\pi^+ p$, which is a typical three-pion
photoproduction reaction. In data analysis  it is easy to reconstruct  $\Delta^{++}$,  so the two-pion production combined with a $\Delta^{++}$ instead of the full three-pion photoproduction will be considered in this work. Moreover,  it is easy to detect in experiment due to the charged final particles. In this work, we will focus on
the decay of a nucleon resonance near 2.1 GeV, especially the interesting 
$N(2120)$. According to the PDG~\cite{prd9}, the resonances with large
branch ratios in the $\Delta\pi$ channel concentrate around 1.7 GeV. Hence, the decays of nucleon resonances near 2.1 GeV to a nucleon resonance near 1.7 GeV are focused on in this work. The decay amplitudes will be studied in the constituent quark model. And the coupling constants for the radiative and strong decays will be calculated from the decay amplitudes. With the coupling constants
obtained, the cross section of
pion photoproduction off the neutron target with the $\Delta$ baryon,
i.e., $\gamma n \to\pi^{-}\pi^{-}\Delta^{++}$, will be studied by
using the effective Lagrangian method.

This article is organized as follows. The model used in this work is presented in the next section. The interaction
mechanisms of  $\gamma n \to\pi^{-}\pi^{-}\Delta^{++}$ are presented
and the possible backgrounds are also discussed.  The
coupling constants for pion decays of nucleon resonances are
studied in the constituent quark model under SU(6) $\otimes$ O(3)
symmetry. The numerical results are given in Sec. III. A brief
summary is given in the last section.

\section{Model}
\subsection{Interaction mechanism}

\begin{figure}[h!]
\includegraphics[bb=50 530 620 730,scale=0.35,clip]{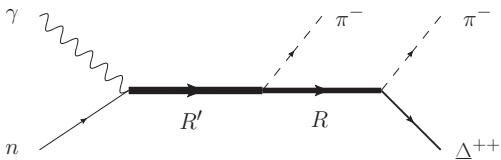}
\caption{Feynman diagrams for the resonance contribution of $\gamma n \to \pi^{-}\pi^{-}\Delta^{++}$ reaction.} \label{feydiagram}
\end{figure}

The interaction mechanisms for the process $\gamma n
\to\pi^{-}\pi^{-}\Delta^{++}$ can be divided into two categories,
namely, resonance contribution and background contribution. The
resonance contribution, which is focused on in this work, includes
mechanisms with the decay of a nucleon resonance to another nucleon
resonance.  As illustrated in Fig.~\ref{feydiagram}, photon $\gamma$
strikes on neutron $n$ and excites it to a nucleon resonance $R'^0$
with higher mass, such as about 2.1 GeV.  Then this
nucleon resonance $R'^0$ decays to a lower resonance $R^+$, which
decays to $\Delta^{++}$ with a pion meson  finally.

To calculate the amplitudes of these diagrams, it is essential to know the Lagrangians for a
vertex for excitation of neutron to nucleon resonance $R'^0$, a vertex
for pion decay of a nucleon resonance $R'^0$ to $R^+$ and a vertex for
pion decay of a nucleon resonance to $\Delta^{++}$. Such Lagrangians have
been constructed in Refs.~\cite{prd10,prd11,prd12} for the resonances
with arbitrary half-integer spin, which are in the forms of

\begin{eqnarray}
{\cal L}_{\gamma N R[\frac{1}{2}^{\pm}]} & = &  \frac{e f_2}{2M_N}\bar{N}\Gamma^{(\mp)}\sigma_{\mu\nu}F^{\mu\nu}R+{\rm H.c.},\\
{\cal L}_{\gamma{}NR[J^{\pm}]}&=&\frac{-i^{n}f_{1}}{(2m_{N})^{n}}\bar{N}^{*}\gamma_{\nu}\partial_{\mu_{2}}...\partial_{\mu_{n}}F_{\mu_{1}\nu}\Gamma^{\pm{}(-1)^{n+1}}R^{\mu_1\mu_2...\mu_{n}}\nonumber\\
                        &+&\frac{i^{n+1}f_2}{(2m_N)^{n+1}}\partial_{\nu}\bar{N}\partial_{\mu_{2}}...\partial_{\mu_{n}}F_{\mu_{1}\nu}\Gamma^{\pm{}(-1)^{n+1}}R^{\mu_1\mu_2...\mu_{n}}\nonumber\\
                        &+&{\rm H.c.},
\end{eqnarray}
where $R^{\mu_1\mu_2...\mu_{n}}$ is the field for the resonance with spin $J=n+1/2$, and
$\Gamma^{(\pm)}=(i\gamma_{5},1)$
{for the different resonance parities}.

The Lagrangian for the strong decay can be written as
\begin{eqnarray}
{\cal L}_{R'[\frac{1}{2}^{\pm}]\pi R[\frac{3}{2}^{\mp}]} & = &  \frac{i g_2}{2M_{\pi}}\partial_{\mu}\pi\bar{R}_1\Gamma^{(\pm)}R+{\rm H.c.},\\
{\cal L}_{R'(J^{\pm})\pi R[\frac{3}{2}^{\mp}]}&=&\frac{i^{2-n}g_{1}}{(2m_{\pi})^{n}}\bar{R^{*}_1}_{\mu1}\gamma_{\nu}\partial_{\nu}\partial_{\mu_{2}}...\partial_{\mu_{n}}\pi\nonumber\\
                       &\times{}&\Gamma^{\pm(-1)^{n}}R^{\mu_1\mu_2...\mu_{n}}\nonumber\\
                        &+&\frac{i^{1-n}g_2}{(2m_{\pi})^{n+1}}\bar{R_1^{*}}_{\alpha}\partial_{\alpha}\partial_{\mu1}\partial_{\mu_{2}}...\partial_{\mu_{n}}\pi\nonumber\\
                        &\times{}&\Gamma^{\pm{}(-1)^{n}}R^{\mu_1\mu_2...\mu_{n}}+{\rm H.c.}.
\end{eqnarray}
The coupling constants will be calculated with the help of the constituent quark model in the next section.
The isospin structure for $\Delta^{*}\pi{}N^{*}$
interaction reads
\begin{align}
\bar{\Delta}^{*}\mathbf{T}_{\frac{3}{2}}\cdot{}\mathbf{\pi{}}N^{*}&=(\sqrt{3}\pi^{-}\Delta^{*++}+\sqrt{2}\pi^{0}\Delta^{*+}+\pi^{+}\Delta^{*0})N^{*+}\nonumber\\
                                                                  &+\pi^{-}\Delta^{*+}N^{*0}+\sqrt{2}\pi^{0}\Delta^{*0}N^{*0}+\sqrt{3}\pi^{+}\Delta^{*-}N^{*0}.
\end{align}

\begin{figure}[h!]
\includegraphics[bb=70 500 780 720,scale=0.32,clip]{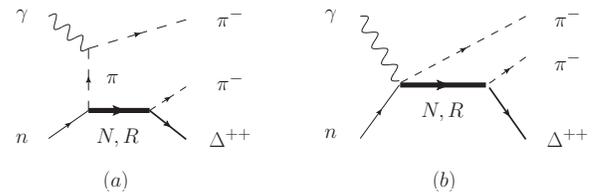}
\caption{Feynman diagrams of background contribution for $\gamma n \to \pi^{-}\pi^{-}\Delta^{++}$ reaction. (a) t-channel, (b) contact term.}
\label{feydiagrams}
\end{figure}

Besides the resonance contribution from cascade decay of nucleon
resonance, there exist other interaction mechanisms, i.\ e.,
background contribution. In experiment, the two-pion production with $t$ channel is often used to research the exotic meson.  Fortunately, it is not involved in the process $\gamma n \to\pi^{-}\pi^{-}\Delta^{++}$
because there does not exist a meson with charge $Q=-2$.   The
mechanisms in Figs.~\ref{feydiagrams}(a) and ~\ref{feydiagrams}(b) with resonances $R^+$
are not considered in this work because the experimentist can remove
such contributions based on the  $\pi^-\Delta^{++}$ invariant mass
spectrum. Hence, the background contribution is mainly from two mechanisms with proton $p$ as shown in Fig. ~\ref{feydiagrams} .  The
contact term is essential to keep the gauge invariance.

The interaction Lagrangians  used in  calculation of background
contributions are~\cite{prd6,He:2012ud},
\begin{eqnarray}
{\cal L}_{\pi NN} & = &  -i g_{\pi NN}\bar{N}\gamma_{5}\gamma_{\mu}{\bm \tau}\cdot\partial^{\mu}{\bm\pi}N,\nonumber\\
{\cal L}_{\pi N\Delta} &=& \frac{g_{\pi N\Delta}}{m_{\pi}}\bar{N}\partial^{\mu}\pi \Delta_{\mu},\nonumber\\
{\cal L}_{\gamma NN} &=& -e\bar{N}\left[Q_{N}A\!\!\!/-\frac{\kappa_N}{4M_{N}}\sigma^{\mu\nu}F^{\mu\nu}\right]N,\nonumber\\
{\cal L}_{\gamma \pi\pi} &=& ie[(\partial^{\mu}\pi^{\dagger})\pi-(\partial^{\mu}\pi)\pi^{\dagger}]A_{\mu},\nonumber\\
{\cal L}_{\gamma \pi NN} &=& -\frac{ie Q_{\pi}g_{\pi NN}}{M_{N}}\bar{N}\gamma_{5}{\bm\tau}\cdot\partial^{\mu}{\bm \pi}A_{\mu}N,
\end{eqnarray}
where $A^\mu$, $\pi$, $N$ and $\Delta$ denote the fields for the
photon, pion meson with mass $m_\pi$, nucleon with mass $m_N$ and
$\Delta$, respectively. Here $Q_h$ is the charge in the unit of
$e=\sqrt{4\pi{\alpha}}$. The anomalous magnetic momentum of the neutron is
$\kappa_{n}=-1.913$~\cite{prd9}. The antisymmetric tensor is defined
by $\sigma_{\mu\nu}=\frac{i}{2}(\gamma_{\mu\nu}-\gamma_{\nu\mu})$ and
$F_{\mu\nu}=\partial^\mu A^\nu-\partial^\nu A^\mu $. The coupling
constant of $g_{\pi N\Delta}$ can be obtained by the experimental
decay width $\Gamma_{\Delta \to \pi N}=116-120$ MeV~\cite{prd9} as
$g_{\pi N\Delta}=2.16\pm 0.02$. The $\pi NN$ coupling constant is
chosen as $g^{2}_{\pi NN}/4\pi=14.4$.

The form factor should be introduced to reflect the internal structure of hadrons. For the $t$-channel exchange in
Fig.~\ref{feydiagrams} (a), form factor has a form
\begin{equation}
F_{\pi}(t)=\frac{\Lambda_{\pi}^2-m_{\pi}^2}{\Lambda_{\pi}^2-t},
\end{equation}
where $t$ is the square of the four-momentum of exchanged $\pi$ meson. The
cut off $\Lambda_{\pi}$ is usually chosen close to 1 GeV.  For
the intermediate $s$ channel theform factor is of the form
\begin{equation}
F(s)=\left(\frac{n\Lambda_{R,N}^4}{n\Lambda_{R,N}^4+(s-M_{R,N}^2)^2}\right)^n,
\end{equation}
where $s$ is the square of the four-momentum of resonances and nucleon, and
the parameter $n=1$ or 2 for the nucleon or resonance intermediate $s$
channel, respectively.  The cut offs will be discussed in Sec.
\ref{results}.

\subsection{Decay of nucleon resonance in the constituent quark model} \label{Formalism}

 To calculate the resonance contribution it is necessary to determine the coupling constants for the decay of nucleon resonances in Eqs. (1)-(4). In Ref.~\cite{Oh:2007jd} the coupling constants was related to the decay amplitudes and this method was applied in the studies about $\Lambda(1520)$ and $\Sigma(1385)$ photoproductions \cite{He:2014gga,He:2012ud,He:2013ksa}.
The decays of a nucleon resonance to a resonance has not been studied in the literature. In this work, the helicity amplitudes of
radiative and strong decay amplitudes of  nucleon resonances will
be calculated within the constituent quark model following the method in Ref. \cite{prd1}.

The decay amplitudes is obtained by making a nonrelativistic reduction of quark-photon/meson interaction. The amplitudes are calculated by sandwiching this interaction between initial and final states under the SU(6) $\otimes$ O(3) symmetry. For example, the definition of the helicity amplitude is as below,
\begin{equation}
	A_\lambda=\frac{1}{\sqrt{2|\bm k|}}\langle N,\lambda-1)|-iH_\gamma|R,\lambda)\rangle
\end{equation}
where $|\bm k|= (M_R^2 - M_n^2)/(2M_R)$ with ${\bm k}$ being the momentum
of photon in the center of mass system of the decaying nucleon
resonance, with $M_R$ and $M_n$ being the masses of resonance and neutron and $\lambda=1/2$ or $3/2$ being the helicity. The explicit form of the interaction and decay amplitude for meson emission have been given explicitly in Ref.~\cite{prd1}.

Here we adopt a non-relativistic constituent quark model to calculate decay amplitudes.  Although this approach results in some loss of "model" accuracy, the non-relativistic constituent quark model can provide a general description of the hadron spectroscopy, which is enough in the current work.  In Ref. \cite{prd108}, it has been found that for light-quark mesons and baryons, the relativistic quark potential model supports the non-relativistic calculations in general.

With the radiative and strong decay amplitudes obtained
the coupling constants $f_1$, $f_2$, $g_1$ and $g_2$
are determined by  the method following Refs.~\cite{Oh:2007jd,He:2014gga,He:2012ud,He:2013ksa}.
Since there are two amplitudes $A_{1/2}$ and $A_{3/2}$ for the
resonances with $J>1/2$, the coupling constants $f_1$ and $f_2$ in
${\cal L}_{Rn\gamma}$ can be extracted from the helicity amplitudes of
the resonance $R$, $A^n_{1/2}$ and
$A^n_{3/2}$. The amplitudes for a state with $J^P=3/2^-$ reads,
\begin{align}
A^n_{1/2}&=\frac{e\sqrt{6}}{12}\sqrt{\frac{|{\bm k}|}{M_{n}M_{R}}}\left[f_1+\frac{f_2}{4M_{n}^2}M_{R}(M_R+M_{n})\right],\\
A^n_{3/2}&=\frac{e\sqrt{2}}{4M_{n}}\sqrt{\frac{|{\bm k}|M_{R}}{M_{n}}}\left[f_1+\frac{f_2}{4M_{n}}(M_R+M_{n})\right].
\end{align}
The coupling constants $g_1$ and $g_2$ can be calculated analogously.
For the nucleon resonances with $J^P=3/2^-$ decaying to a state with
$J^P=5/2^-$ with a pion, the decay amplitudes have a form,
\begin{align}
	{{A}}^\pi_{3/2}&=-\frac{\sqrt{3}g_1}{(2M_{\pi})^2}{\cal{B}}{\cal{F}},\\
	{A}^\pi_{1/2}&=-\frac{g_1}{\sqrt{2}(2M_{\pi})^2}\left(\frac{2|{\bm p_{\pi}}|^2}{M_{R}(M_{R}+E_{R})}+3\right){\cal{B}}{\cal{F}}\nonumber\\
                               &+\frac{\sqrt{2}g_2}{(2M_{\pi})^3}\frac{|{\bm p_{\pi}}|(E_{R}+M_{R})}{\sqrt{5M_{R}(E_{R}+M_{R})}}{\cal{B}}^2,
\end{align}
where ${\cal{B}}=\left(\frac{E_{\pi}}{M_{R}}+\frac{|{\bm p_{\pi}}|^2}{M_{R}(M_{R}+E_{R})}+1\right)|{\bm p_{\pi}}|$ and ${\cal{F}}=\frac{E_{\pi}(E_{R}+M_{R})+|{\bm p_{\pi}}|^2}{\sqrt{5M_{R}(E_{R}+M_{R})}}$ with ${\bm p_{\pi}}$ and $E_{\pi}$
being momentum and energy of $\pi^{-}$. $M_{R}$ and $E_{R}$ are mass and energy for a resonance $R$, respectively.

For a nucleon resonance with $5/2^-$ decaying to $\Delta\pi$, we have
\begin{align}
	{A}^\pi_{3/2}&=\frac{i\sqrt{3}g_1}{(2M_{\pi})^2}{\cal{D}},\\
	{A}^\pi_{1/2}&=\frac{ig_1}{\sqrt{2}(2M_{\pi})^2}\left(\frac{2|{\bm p_{\pi}}|^2}{M_{\Delta}(M_{\Delta}+E_{\Delta})}+1\right){\cal{D}}\nonumber\\
                               &+\frac{i\sqrt{2}g_2}{(2M_{\pi})^3}\frac{|{\bm p_{\pi}}|^3}{\sqrt{5M_{\Delta}(E_{\Delta}+M_{\Delta})}}{\cal{B}^{'}},
\end{align}
where ${\cal{B}^{'}}=\left(\frac{E_{\pi}}{M_{\Delta}}+\frac{|{\bm p_{\pi}}|^2}{M_{\Delta}(M_{\Delta}+E_{\Delta})}+1\right)|{\bm p_{\pi}}|$ and ${\cal{D}}=\frac{(E_{\Delta}+M_{\Delta}+E_{\pi})}{\sqrt{5M_{\Delta}(E_{\Delta}+M_{\Delta})}}|{\bm p_{\pi}}|^2$. $M_{\Delta}$ and $E_{\Delta}$ are mass and energy for $\Delta$, respectively.

\section{results}

The decays of nucleon resonance $R'$ near 2.1 GeV to nucleon resonance $R$ 
near 1.7 GeV  are focused on in the current work.  The radiative decay
widths for $R'^0\to n\gamma$ and the strong decay widths for $R^+\to
\pi^-\Delta^{++}$ can be used to preliminarily select the nucleon
resonances $R$ and $R'$ which are important in the resonance
contribution. With the dominant channels selected, the cross sections of process $\gamma n \to\pi^{-}\pi^{-}\Delta^{++}$ will be calculated and the possibility of the experimental study will be discussed.

\subsection{Decay amplitudes and coupling constants}

The numerical results of helicity amplitudes for radiative and pion decays are
listed in Tables~\ref{Nn} and \ref{ND} and the  values suggested for three or four star resonances listed in
the PDG are also presented for comparison.  The resonances, which are not
in the mass range concerned in this work, are excluded based on
Ref.~\cite{prd17}.
 The state under
SU(6) $\otimes$ O(3) in constituent quark model is labeled by the
notation $X(^{2S+1}L_{\pi})J^{P}$ in which $X=N$ or $\Delta$, {$S$ is
the quark total spin}, $L$ is the orbital angular momentum, $\pi$ is the
permutational symmetry(symmetric,mixed,antisymmetric)of the spatial
wave function, and $J^{P}$ is the total angular momentum and parity of
the state.

\renewcommand\tabcolsep{0.27cm}
\renewcommand{\arraystretch}{1.6}
\begin{table}[h!]
\caption{The amplitudes and decay width for $R'^0\to n\gamma$. The experimental
values are given under the theoretical values for three or four star
resonances listed by the PDG ~\cite{prd9}. \label{Nn}}
\begin{tabular}{lrrr}
\hline\hline
State&                       $A_{3/2}^{n}$ [GeV$^{-1/2}$]   &
$A_{1/2}^{n}$ [GeV$^{-1/2}$]  &             $\Gamma$ [MeV] \\\hline
$N(^{4}D_{M})\frac{7}{2}^{+}$      & $-0.023$                         &         -0.018                           &             $0.020$ \\
$N(^{2}P_{M})\frac{3}{2}^{-}$      & $-i0.057$                        &         $i0.015$                         &             $0.180$\\
$N(^{4}P_{M})\frac{3}{2}^{-}$     & $i0.079$                         &         $i0.015$                         &             $0.328$\\
$N(^{4}D_{M})\frac{3}{2}^{+}$     & $0.018$                          &         $-0.010$                         &             $0.002$ \\
$N(1900)$                         &$-0.010\pm0.004$ &
$-0.011\pm0.007$\\
$N(^{4}P_{M})\frac{5}{2}^{-}$    & $-i0.063$                        &         $-i0.045$                        &             $0.080$ \\
$\Delta(^{4}D_{S})\frac{7}{2}^{+}$  & $-0.045$                         &        $-0.035$                           &             $0.071$ \\
$\Delta(1950)$                         &$-0.076\pm0.012$ &
$-0.097\pm0.010$\\
$\Delta(^{2}P_{M})\frac{3}{2}^{-}$  &$i0.073$                          &        $i0.088$                           &             $0.561$\\
$\Delta(^{4}D_{S})\frac{3}{2}^{+}$  &$-0.036$                          &        $0.021$                           &              $0.070$\\
$\Delta(1920)$                        &$-0.040\pm0.014$ & $0.023\pm0.017$\\
$\Delta(^{4}D_{S})\frac{1}{2}^{+}$  & $--$                               &        $-0.021$                          &              $0.035$\\
$\Delta(1910)$                     &   $--$&$0.020\pm0.010$ &\\
$\Delta(^{4}D_{S})\frac{5}{2}^{+}$ & $0.057$                          &        $0.013$                           &              $0.095$\\
$\Delta(1905)$                         &$0.022\pm0.005$ &
$-0.045\pm0.010$\\
$\Delta(^{2}P_{M})\frac{1}{2}^{-}$  & $--$                               &        $i0.075$                          &              $0.077$\\
\hline\hline
\end{tabular}
\end{table}

\renewcommand\tabcolsep{0.17cm}
\begin{table}[h!]
\caption{The amplitudes and  decay width for $R^+\to \pi\Delta$. The
	suggested values by the PDG are listed in the last column~\cite{prd9}.\label{ND}}
\begin{tabular}{ccrrrr}
\hline\hline
$State$&                           ${\rm PDG}$&
$A^\pi_{3/2}$&         $A^\pi_{1/2}$&
$\Gamma$ [MeV] & Exp.~\cite{prd9}\\\hline
$N(^{4}P_{M})\frac{1}{2}^{-}$&     $N(1650)$&       $--$&
$i0.501$&                                $14.2$ & 0-34\\
$N(^{4}P_{M})\frac{5}{2}^{-}$&     $N(1675)$&           $-i1.182$&                        $-i0.847$&
$43.4$ & 65-90 \\
$N(^{2}D_{S})\frac{5}{2}^{+}$&     $N(1680)$&           $-0.533$&
$0.000$&
$4.0$ &6-20 \\
$N(^{2}S_{M})\frac{1}{2}^{+}$&     $N(1710)$&           $--$&
$-0.067$&                                $0.3$ &8-40\\
$N(^{2}D_{S})\frac{3}{2}^{+}$&     $N(1720)$&           $-0.282$&                         $-0.208$&
$4.2$\\
\hline \hline
\end{tabular}
\end{table}

 One can find that the values obtained in the constituent quark model
 without mixing effect are comparable with the suggested values by the
 PDG~\cite{prd9}.  Among the resonances near 2.1 GeV, the radiative
 decay widths of states $N(^4P_M)\frac{3}{2}^{-}$,
 $N(^2P_M)\frac{3}{2}^{-}$ and $\Delta(^2P_M)\frac{3}{2}^{-}$  are
 much larger than the decay widths of other resonances. And among the
 resonances near 1.7 GeV, state $N(^4P_M)\frac{1}{2}^{-}$, which
 corresponds to the $N(1650)$, and state $N(^4P_M)\frac{5}{2}^{-}$,
 which corresponds to the $N(1675)$, have relatively large decay width
 in $\pi^-\Delta^{++}$ channel.

The helicity amplitudes for $R'^0\to R^+\pi^{-}$  can be calculated
analogously, which are tabulated in Table ~\ref{BBp}. Here only the
nucleon resonances selected in the above step are considered. In the
channel $N(^{2}P_{M}){3\over2}^-\to N(^{4}P_{M}){5\over2}^-\pi^{-}$, a
much large decay width can be found, which is expected to dominate in
the total process.

\renewcommand\tabcolsep{0.27cm}
\begin{table}[h!]
\caption{The amplitudes and decay width for $R'^0\to R^+\pi^{-}$. \label{BBp}}
\begin{tabular}{crrr}
\hline\hline
Channel&                                           $A^\pi_{3/2}$&
$A^\pi_{1/2}$&             $\Gamma$ [MeV]  \\\hline
$N(^{4}P_{M}){3\over2}^-\to N(^{4}P_{M}){1\over2}^-\pi^{-}$&
$--$&                       $1.520$&                               $29.40$\\
$N(^{4}P_{M}){3\over2}^-\to N(^{4}P_{M}){5\over2}^-\pi^{-}$&      $0.856$&                       $1.091$&                               $23.46$\\
$N(^{2}P_{M}){3\over2}^-\to N(^{4}P_{M}){1\over2}^-\pi^{-}$&
$--$&                       $1.373$&                               $23.99$\\
$N(^{2}P_{M}){3\over2}^-\to N(^{4}P_{M}){5\over2}^-\pi^{-}$&     $-2.869$&                       $-3.134$&                              $220.18$\\
$\Delta(^{2}P_{M}){3\over2}^-\to N(^{4}P_{M}){5\over2}^-\pi^{-}$&$-1.006$&                       $-0.929$&                              $13.68$\\
$\Delta(^{2}P_{M}){3\over2}^-\to N(^{4}P_{M}){1\over2}^-\pi^{-}$&
$--$&                       $0.501$&                               $2.03$\\
\hline \hline
\end{tabular}
\end{table}

Based on the analysis above, two channels $\gamma n\to
N(^{4}P_{M}){3\over2}^-\to
N(^{4}P_{M}){1\over2}^-\pi^{-}\to\Delta^{++}\pi^-\pi^-$ and $\gamma
n\to N(^{2}P_{M}){3\over2}^-\to
N(^{4}P_{M}){5\over2}^-\pi^{-}\to\Delta^{++}\pi^-\pi^-$ will be
considered in the calculation of the cross section and larger
contributions are expected from the later channel. The coupling
constants $f_1$ and $f_2$ for radiative decay and $g_1$ and $g_2$ for
$\pi^-$ decay  can be calculated with the formalism given in the
previous section.  The explicit values of the coupling constants for
the selected channels are listed in Table~\ref{Table: cc}.
As shown in Eq.(3), there is only one coupling constant $g_2$ for the decay
of a state with spin 3/2 into one with spin 1/2. The analogous can be
found for the  decay of a state with spin 1/2  into
one with spin 3/2.

\renewcommand\tabcolsep{0.57cm}
\begin{table}[h!]
\caption{The coupling constants $f_1$ and $f_2$ for radiative decay and $g_1$ and $g_2$ for $\pi^-$ decay.\label{Table: cc}}
\begin{tabular}{lrr}
\hline\hline
 Channel   &                                              $f_1$   &        $f_2$  \\\hline
$N(^{2}P_{M})\frac{3}{2}^{-}\to{}\gamma{}n$      & $-0.938$ &      0.705 \\
$N(^{4}P_{M})\frac{3}{2}^{-}\to{}\gamma{}n$      & $0.593$  &
-0.119\\\hline\hline
 Channel   &                                               $g_1$  &        $g_2$ \\\hline
$N(^{4}P_{M})\frac{5}{2}^{-}\to{}\Delta^{++}   \pi^{-}$&   $-0.074$&   $6.664$\\
$N(^{4}P_{M})\frac{1}{2}^{-}\to{}\Delta^{++}   \pi^{-}$&   $--$&   $0.101$\\
$N(^{2}P_{M})\frac{3}{2}^{-}\to{}N(^{4}P_{M})\frac{5}{2}^{-}\pi^{-}$&   $0.130$&   $6.929$\\
$N(^{4}P_{M})\frac{3}{2}^{-}\to{}N(^{4}P_{M})\frac{1}{2}^{-}\pi^{-}$&
$--$&   $3.733$\\
\hline\hline
\end{tabular}
\end{table}

\subsection{Cross sections and Dalitz plot}\label{results}

\begin{figure}[h!]
\includegraphics[bb=40 30 800 600,scale=0.35,clip]{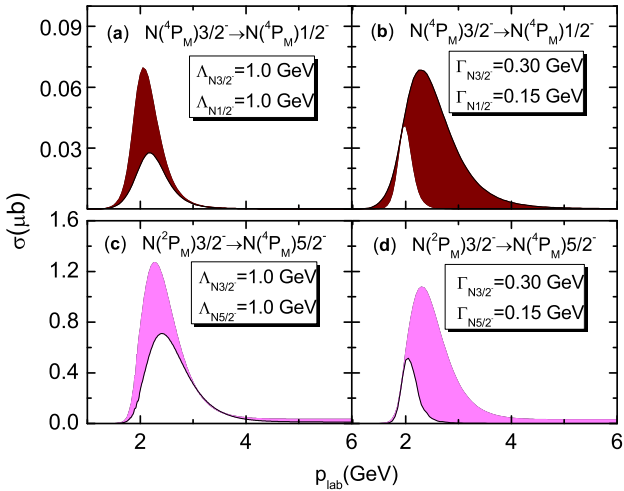}
\caption{(Color online) The total cross section for $\gamma n\to\pi^{-}\pi^{-}\Delta^{++}$  from the resonance contributions versus the photon-beam energy $P_{lab}$. The panels (a) and (c) are obtained with variation of widths of higher resonance from 0.25 to 0.45 GeV with the cut off $\Lambda=1.0$ GeV.  The panels (b) and (d) with variation of the cut off of higher resonance from 0.7 to 1.3 GeV  with width $\Gamma=0.3$ and 0.15 GeV for the higher and lower resonances,respectively. }  \label{Fig:tcsnrhopi}
\end{figure}

With the Lagrangians and the coupling constants obtained in the
previous section, the cross section versus {the photon-beam energy $P_{lab}$}
 is calculated with the help of code FOWL in the CERNLIB
program. The cross sections from the resonance contribution with the
variation of the beam energies $P_{lab}$ is presented in
Fig.~\ref{Fig:tcsnrhopi}.
\begin{figure}[h!]
\includegraphics[bb=20 30 650 600,scale=0.4,clip]{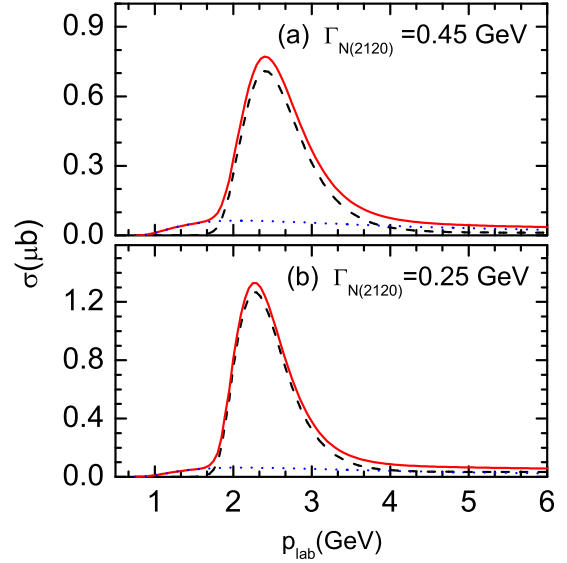}
\caption{(Color online) The total cross section  for
$\gamma n\to\pi^{-}\pi^{-}\Delta^{++}$ with width $\Gamma_{N(2120)}=0.45$ and $0.25$ GeV and cut off $\Lambda=1$ GeV.
The solid (red), dashed (black) and dotted (blue) lines are for full model, resonance contribution and background contribution.
} ~\label{Fig:gammandpp}
\end{figure}

The decay widths in Tables ~\ref{Nn}-\ref{BBp} suggest
that the most important contribution is from the mechanism with decay
$N(^2P_M)\frac{3}{2}^-\to N(^4P_M)\frac{5}{2}^-$. It is confirmed by
the numerical results as shown in Fig.~\ref{Fig:tcsnrhopi} where both
results with decay $N(^2P_M)\frac{3}{2}^-\to
N(^4P_M)\frac{5}{2}^-$ and results of the second most important
mechanism with decay $N(^4P_M)\frac{3}{2}^-\to N(^4P_M)\frac{1}{2}^-$
are illustrated. The uncertainties arising from the total decay widths
and cut offs are also considered. The results with variation of widths
from 250 to 450 MeV are shown in Fig.~\ref{Fig:tcsnrhopi}(a) and \ref{Fig:tcsnrhopi}(c)
and the results with variation of cut off  from 0.7 to 1.3 GeV is
shown in Fig.~\ref{Fig:tcsnrhopi}(b) and \ref{Fig:tcsnrhopi}(d). After these
uncertainties are considered, one can find that the contribution from the mechanism
with decay $N(^2P_M)\frac{3}{2}^-\to N(^4P_M)\frac{5}{2}^-$ is more
important than these with other cascade decays. In the constituent quark
model, the state $N(^2P_M)\frac{3}{2}^-$ corresponds to the $N(2120)$
and the state $N(^4P_M)\frac{5}{2}^-$ corresponds to the $N(1675)$.
Hence, one can say that in  the process $\gamma
n\to\pi^{-}\pi^{-}\Delta^{++}$ the contribution from the mechanism
with the decay of the $N(2120)$ to the $N(1675)$ is dominant. And the
cross section is  at order of 1 $\mu$b.

\begin{figure}[h!]
\includegraphics[bb=30 280 550 750,scale=0.43,clip]{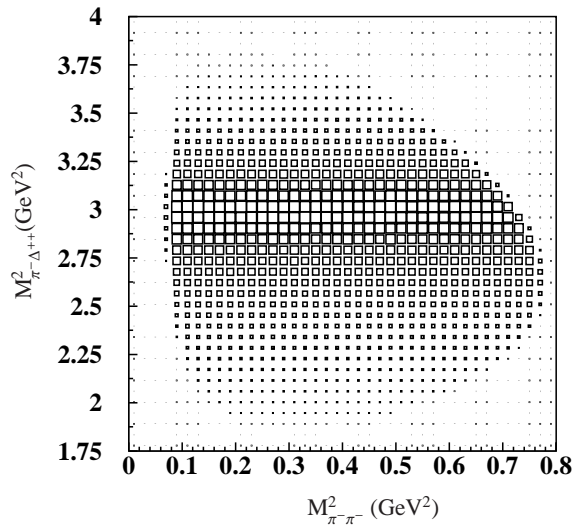}
\caption{The Dalitz plot for the process $\gamma n \to\pi^{-}\pi^{-}\Delta^{++}$.
} ~\label{Fig:iddp}
\end{figure}

In Fig.~\ref{Fig:gammandpp} the total cross sections with background
contribution with $\Lambda=1.0$ GeV and $\Gamma_{N(2120)}=0.45$ and
$0.25$ GeV  are  presented. The Dalitz plot is also shown in
Fig.~\ref{Fig:iddp} for reference. The resonance contribution is
significant compared with the background.

\section{Summary}

In the present work,  the possibility to reseach the decay of the
$N(2120)$ to the $N(1675)$ in photoproduction $\gamma n
\to\pi^{-}\pi^{-}\Delta^{++}$ is explored.  The decay amplitudes  of nucleon resonances are calculated in the constituent quark model.  The results suggest that a mechanism with decay of the $N(2120)$  to the $N(1675)$ is dominant in the process $\gamma n\to\pi^{-}\pi^{-}\Delta^{++}$. Other contributions  from other nucleon  resonances are relatively small and can be neglected.

Numerical calculations for the cross section of the process $\gamma
n\to\pi^{-}\pi^{-}\Delta^{++}$ are performed with the effective
Lagrangians and the coupling constants obtained from the decay amplitudes. The results confirm that the mechanism with decay of the
$N(2120)$ to the $N(1675)$ is dominant and the background
contributions are very small. The total cross section of  process
$\gamma n\to\pi^{-}\pi^{-}\Delta^{++}$ is at order 1$\mu$b, which is
large enough to be studied in JLab.

Due to the small background contribution, the peak near 2.1 GeV as shown in Fig. \ref{Fig:gammandpp} is an obvious signal in experiment to search for the $N(2120)$.
 According to our calculation, selection of the $N(1675)$ after  reconstruction from $\Delta\pi$ will make the signal for $N(2120)$ clearer because of the large decay width of  $N(2120)\to N(1675) \pi$.  This
interesting exploration of nucleon resonance near 2.1 GeV can be
performed in the future experiments of CLAS12@JLab.

\section*{Acknowledgments}

This project was partially supported by the Major State
Basic Research Development Program in China (No. 2014CB845400),
the National Natural Science
Foundation of China (Grants No. 11275235, No. 11035006, No.11175220)
and the Chinese Academy of Sciences (the Knowledge Innovation
Project under Grant No. KJCX2-EW-N01, Century program under Grant No. Y101020BR0).

\end{document}